\documentclass[12pt,preprint]{aastex}

\shorttitle{Angular Clustering of SWIRE Galaxies}
\shortauthors{Oliver et al.}

\begin{document}

\title{Angular Clustering of Galaxies at 3.6~\micron\ from the SWIRE survey}

\author{Seb Oliver and Ian Waddington}
\affil{Astronomy Centre, Department of Physics \& Astronomy,
  University of Sussex, Brighton, BN1 9QH, UK}
\email{S.Oliver@Sussex.ac.uk}

\author{Eduardo Gonzalez-Solares\altaffilmark{1}} 
\affil{University of Cambridge, Institute of Astronomy, The
  Observatories, Madingley Road, Cambridge, CB3 0HA, UK}

\author{Jason Surace\altaffilmark{2}, Fan Fang\altaffilmark{2}, Dave
  Shupe\altaffilmark{2}, Tom Jarrett, Carol Lonsdale, Cong (Kevin)
  Xu and Duncan Farrah} 
\affil{Infrared Processing \& Analysis Center, California Institute of
  Technology, MS 100-22, Pasadena, CA 91125, USA}

\author{Malcolm Salaman}
\affil{Astronomy Centre, Department of Physics \& Astronomy, 
  University of Sussex, Brighton, BN1 9QH, UK}

\author{Michael Rowan-Robinson}
\affil{Astrophysics Group, Blackett Loboratory, Imperial College,
  Prince Consort Road, London, SW7 2BW, UK}

\and

\author{Brian Siana and H. E. (Gene) Smith, }
\affil{Center for Astrophysics \& Space Sciences, University of
  California San Diego, La Jolla, CA 92093-0424, USA}


\altaffiltext{1}{Formerly at University of Sussex}
\altaffiltext{2}{Spitzer Science Center, California Institute of
  Technology, MS 220-6, Pasadena, CA 91125, USA}

\begin{abstract}
We present the first analysis of large-scale clustering from the
Spitzer Wide-area InfraRed Extragalactic legacy survey: SWIRE.  We
compute the angular correlation function of galaxies selected to have
3.6~\micron\ fluxes brighter than 32~$\mu$Jy in three fields totaling
two square degrees in area. In each field we detect clustering with a
high level of significance. The amplitude and slope of the correlation
function is consistent between the three fields and is modeled as
$w(\theta)=A\theta^{1-\gamma}$ with
$A=(0.6\pm0.3)\times10^{-3},\gamma=2.03\pm0.10$.  With a fixed slope
of $\gamma=1.8$, we obtain an amplitude of
$A=(1.7\pm0.1)\times10^{-3}$. Assuming an equivalent depth of
$K\approx18.7$~mag we find our errors are smaller but our results are
consistent with existing clustering measurements in $K$-band surveys
and with stable clustering models.  We estimate our median redshift
$z\simeq 0.75$ and this allows us to obtain an estimate of the
three-dimensional correlation function $\xi(r)$, for which we find
$r_0=4.4\pm0.1 h^{-1}$~Mpc.

\end{abstract}


\section{Introduction}

The Spitzer Wide-area InfraRed Extragalactic legacy survey, SWIRE
\citep{lonsdale2003,lonsdale2004}, has recently begun observations.
The survey was designed to dramatically enhance our understanding of
galaxy evolution. We will study the history of star-formation, the
assembly of stellar mass, the nature and impact of accretion processes
in active nuclei, and the influence of environment on these processes
at all scales.  The survey will detect around two million galaxies at
infrared wavelengths ranging from 3.6~\micron\ to 160~\micron, over
50~square degrees.

The analysis of the clustering of galaxies has been an important tool
in cosmology for many years \citep[e.g.,][]{peebles80}.  Originally
galaxies were assumed to trace the mass density field (modulo some
``bias'' factor) and their clustering was used to constrain
cosmological models.  For example, the angular correlation function of
the APM galaxy survey was able to rule out the once standard cold dark
matter model \citep{maddox90}. Today the cosmological models are
usually constrained by observations of the cosmic microwave background
(e.g. \citealt{wmap}) and conventional models of the evolution of
structure under gravity can then provide us with estimates of the
statistical properties of the mass density field. Hence studies of
galaxy clustering can now be used to understand the relationship
between galaxy formation and the mass field, i.e.\ the galaxy bias
(e.g. \citealt{benson}).

In this paper we take the first step towards an understanding of the
clustering of the SWIRE sources, by measuring the angular correlation
function of those galaxies that are detected at 3.6~\micron\ ($L$
band) with the Infrared Array Camera (IRAC).  For galaxies at
$z\approx 0.6-0.7$ the 3.6~\micron\ band probes the rest-frame
$K$-band and so their clustering can be compared directly with that
from local $K$-band surveys such as the Two Micron All Sky Survey
\citep[2MASS;][]{maller03}.  At this wavelength, the emission is
dominated by older stellar populations so these galaxies are tracing
the sites of star-formation in the distant past and we are probing the
bias at even earlier epochs.

\section{Catalogs and Sample Selection}
\subsection{Catalogs}

The catalogs that we are using are all the ones that were available in
March 2004, i.e.\ our validation ``tile'' in the Lockman hole region
(0.4~deg$^2$) and two tiles in the ELAIS N1 region (each 0.8~deg$^2$).
The data are available from the Spitzer Science Archive as programs
PROGID=142 and PROGID=185 respectively.  The data processing will be
described in detail in a future paper \citep{swiredata}.  The
5$\sigma$ limit of the 3.6 $\mu$m catalogs is $S_{3.6}=3.7$~$\mu$Jy
\citep{lonsdale2004}; in the analysis that follows we consider only
those galaxies that have $S_{3.6}\ge 32$~$\mu$Jy, a flux-limited
sample with typically very high signal-to-noise ratio ($>40$).

\subsection{Star/galaxy Separation}\label{sec:stargal}

Contaminating stars will reduce the galaxy clustering signal, and so
we have investigated two procedures for separating stars and galaxies.
The first method uses our supporting optical data in the central
0.3~square degrees of the Lockman tile.  We select objects with
optical counterparts and detections in at least four bands (U,
$g^\prime$, $r^\prime$, $i^\prime$, 3.6~\micron, 4.5~\micron).  From
these, we reject objects that are morphologically classified as
stellar and are brighter than $r^\prime<23$~mag (below this limit, the
star/galaxy separation becomes unreliable).  We also reject those
objects which have good optical/infrared data, but for which no galaxy
or AGN template spectrum provides a good fit \citep{mrr04}.  This
method provided us with 872 stars and 2115 galaxies, with 727 IRAC
sources rejected. The stellar number count models of \citet{jarrett}
predict 620 sources, therefore we expect the stellar contamination in
our catalog to be small.

In the second method we use the IRAC color from 3.6 to 4.5~\micron:
$C_{12}=\log (S_{3.6}/S_{4.5})$.  We model this color above
$S_{3.6}>500$~$\mu$Jy as a Gaussian (assuming these to be stars, but
fitting only to the positive half of the distribution to avoid
residual contamination from galaxies).  We find a mean
$\bar{C}_{12}=0.235$ and $\sigma_{c_{12}}=0.014$.  We then apply a
three-sigma cut and reject stars with $0.194\le C_{12}\le0.276$. With
this color criterion and the 3.6~\micron\ flux limit, the
4.5~\micron\ completeness limit of $S_{4.5}\ge 18$~$\mu$Jy has no
impact on the identification of the stars.  In addition, we exclude
all objects associated with stars from the 2MASS catalog
\citep{2mass}.  The number of star rejected is shown in
Table~\ref{tab:wtheta}.  The stellar count model \citep{jarrett}
predicts 2074 stars per square degree, thus this method also excludes
around 2000 galaxies per square degree with stellar colors.  These are
a minor fraction of our sample ($<20$ per cent) and this will not bias
our results, so long as the excluded galaxies do not cluster
differently to the remaining sample.  Since we include galaxies with
both redder and bluer colors, this seems a reasonable assumption.

We compared the two different methods of star/galaxy separation in the
0.3~square degrees of the Lockman field where both methods could be
applied.  In this region, the optical selection yielded 872 stars and
the infrared selection method yielded 1100 stars.  Of these, 450
sources are common to both lists, making them the most reliable
stellar identifications.  However, these common sources are not a
complete list of the stars -- the stellar count model \citep{jarrett}
predicts 620 sources in this field, leaving around 170 stars (27 per
cent) that were not selected by both methods concurrently.  Further,
if we consider the 2MASS point source catalog to be a second reliable
list of stellar identifications, then we find that 96 of these 2MASS
stars (23~per cent) were not identified as such using the optical
selection method.  Combining these statistics gives an estimate of the
stellar contamination in the galaxy catalogs of $<5$--8~per cent for
the Lockman optical/IRAC catalog, and $<3$~per cent for the three
IRAC-only catalogs.

\subsection{Selection function}

We have adopted a simple but conservative selection function. Our high
flux limit ($S_{32\mu{\rm m}} \ge 32$~$\mu$Jy, typically
signal-to-noise $>40$) means that even in regions of higher than
average noise (low coverage) we will still have reasonable
signal-to-noise and be well above the completeness level, thus our
selection function is uniform.  A bright star can cause artifacts that
will affect our source detection, however, so we excluded regions
around bright 2MASS stars, and also regions near the boundaries of the
survey fields.

\section{The angular correlation function}

We calculated the angular correlation function
$w(\theta)=A\theta^{1-\gamma}$ following the same techniques as
\citet{eduardolss}, using the \citet{landyszalay} estimator.  We
applied corrections for the finite survey area (the integral
constraint) and the stellar contamination, following the method of,
for example, \citet{roche99}.

\begin{figure}
\epsscale{1.}
\plotone{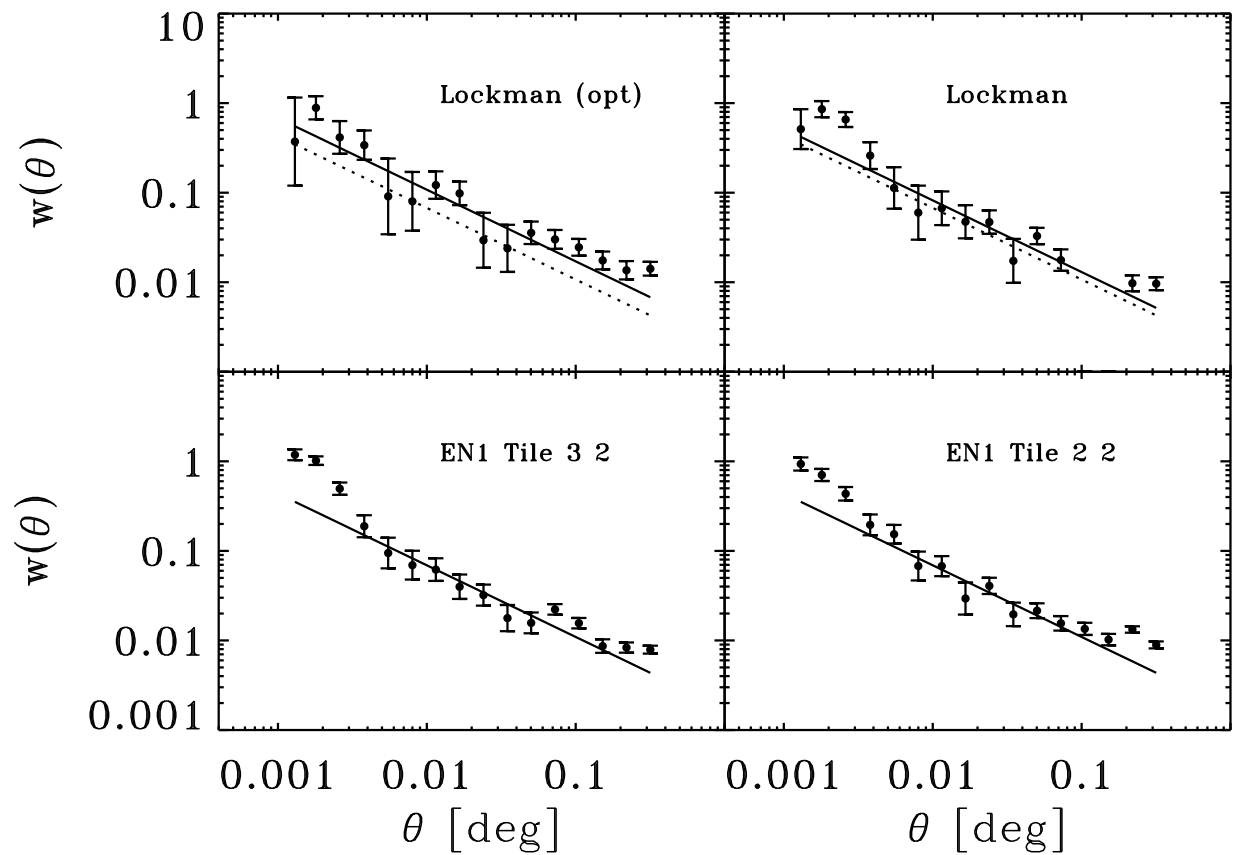}
\epsscale{1.}
\caption{The angular correlation function $w(\theta)$ in each of our
  fields. The best-fitting power law with fixed slope is shown as a
  solid line. This power law is the same for both ELAIS N1 tiles,
  $w(\theta)=0.0017\, (\theta/{\rm deg})^{-0.8}$, and this is plotted
  as a dotted line in the other panels for reference.
}\label{fig:wtheta}
\end{figure}

We calculated $w(\theta)$ for each of our three independent fields:
Lockman, ELAIS N1 tile 2\_2 and ELAIS N1 tile 3\_2. We have made a
seperate measurement in the center of the Lockman field with deep
optical coverage, using a different star/galaxy classification as
described in Section \ref{sec:stargal}.  Each of these correlation
functions are plotted in Figure~\ref{fig:wtheta} and tabulated in
Table~\ref{tab:wtheta}.  We note that, as with any angular correlation
analysis, the data points are not independent.  We fit the model to
the data over $\theta>0.003$~deg (11\arcsec) -- on smaller scales the
correlation function clearly departs from a power law, indicating an
excess of close pairs relative to the clustering on larger scales.
This excess is most-likely due to interacting/merging galaxies
\citep{roche99}, but our large source detection aperture (6\arcsec)
limits our ability to investigate this so we will explore this in more
detail in a future paper. The strength of clustering in the Lockman
optical field appears to higher than the other fields; this might be
because the additional optical selection criteria reduce the effective
depth of the field and shallower surveys will always have stronger
angular clustering.  The best-fitting parameters to the three combined
samples (excluding the Lockman optical dataset) are:
$A=(1.7\pm0.1)\times10^{-3}$ for a fixed $\gamma=1.8$; and
$A=(0.6\pm0.3)\times10^{-3}$ with $\gamma=2.03\pm0.10$ for a free fit.

\subsection{Comparison with $K$-band surveys}

The sources detected in the 3.6~\micron\ band will be similar in
nature to sources detected in a $K$-band survey, as both sample the
old stellar populations.  In the absence of $K$-band data in our
fields, we used simulated catalogs from \citet{xu2003} to estimate the
effective $K$-band limit in three different ways, giving answers
ranging from $K\le18.1$~mag to $K\le19.3$~mag.  Although this model
over-predicts the 3.6~\micron\ number counts \citep{lonsdale2004} we
only need the colours and not the overall normalisation to be correct.

Firstly we determined the median $L-K$ color of galaxies with
$S_{3.6}\ge32$~$\mu$Jy to be 1.1; our limiting flux thus translates to
$K\le18.6$~mag.  Secondly we examined the $K$-band counts of a
$S_{3.6}\ge32$~$\mu$Jy simulated catalog and estimated a completeness
limit of $K\le18.1$~mag (N.B. the overall normalisation of the count
model does not affect this result).  The parameter of interest is not
really the $K$ magnitude but redshift; so for our third method we
constructed $K$ magnitude limited samples with varied limits and found
that a magnitude limit of $K\le19.3$ gives the same median redshift as
we predict for our sample ($z=0.75$, see Section \ref{sec:nz} below).
Thus we define our effective $K$-band limit to be $K=18.7$~mag,
recognizing that there is some uncertainty in this value by up to
$\pm0.6$~mag.

\begin{figure}
\epsscale{0.8}

\plotone{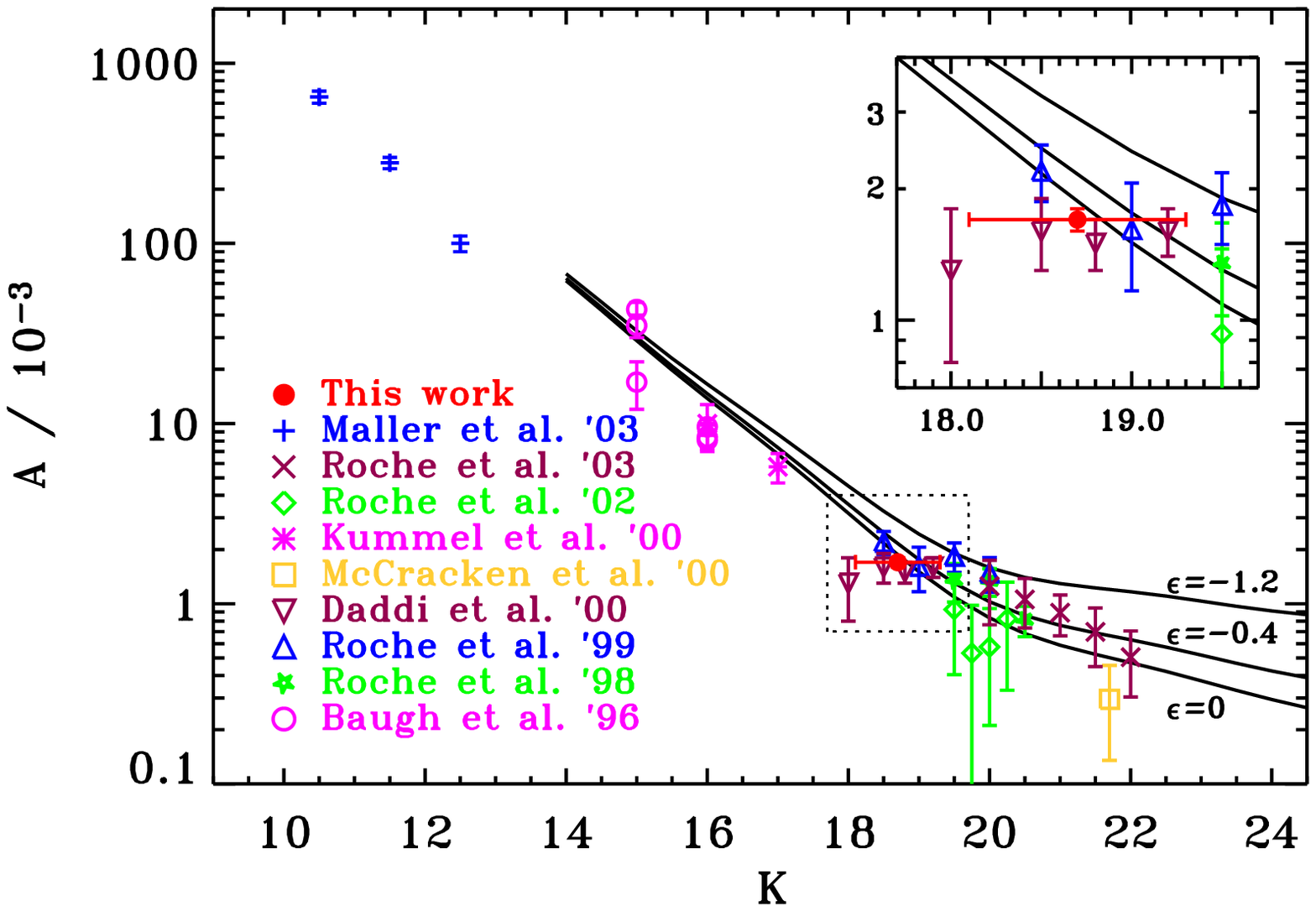}
\epsscale{1.}

\caption{Amplitude $A$ of the angular correlation function $w(\theta)$
  as a function of $K$-band magnitude.  We have plotted our survey at
  an effective $K=18.7$ mag.  Existing data comes from
  \protect\citet{baugh96,roche98,roche99,daddi00,
    mccraken00,kummel00,roche02,roche03,maller03}. All data have fixed
  $\gamma=1.8$ except for those of \protect\citet{maller03} who found
  $\gamma\approx1.76$, depending on scale.  The lines are the evolving
  models of \protect\citet{roche03}, with stable ($\epsilon=0$),
  co-moving $\epsilon=-1.2$ and intermediate $\epsilon=-0.4$
  clustering.  An expanded view of our new measurement is shown in the
  inset.}\label{fig:comp}

\end{figure}

In Figure~\ref{fig:comp} we compare our amplitude with that from
various $K$-band surveys. Our errors are much smaller than existing
measurements in these range but there is a very good agreement, the
main issue is the uncertainty in how to compare the $K$- and $L$-band
limits.

\citet{roche03} have modeled the clustering of $K$-band sources with
the auto-correlation function evolving as
$$\xi(r,z)=\left(r/r_0\right)^{-\gamma}\left(1+z\right)^{-(3+\epsilon)}$$
Setting the local $r_0=5.85h^{-1}$~Mpc \citep{cabanac} and using their
own model for the redshift distribution, they estimate the angular
correlation function as shown in Figure~\ref{fig:comp}. We find good
agreement with their stable clustering, $\epsilon=0$, model.
Interestingly their model appears to over-predict the strength of the
2MASS clustering \citep{maller03}, though these data have been plotted
with the slightly shallower best-fit slopes $\gamma\approx1.76$.

\subsection{Estimation of 3D clustering}\label{sec:nz}

We can use our two-dimensional clustering measurement to infer
the three-dimensional clustering statistics. To do this we need
to know the shape of the redshift distribution $dN/dz(z)$.  In the
absence of spectroscopic redshifts or photometric redshifts for all
our sources we use the model of \citet{xu2003} to estimate the
redshift distribution.

\begin{figure}
\plotone{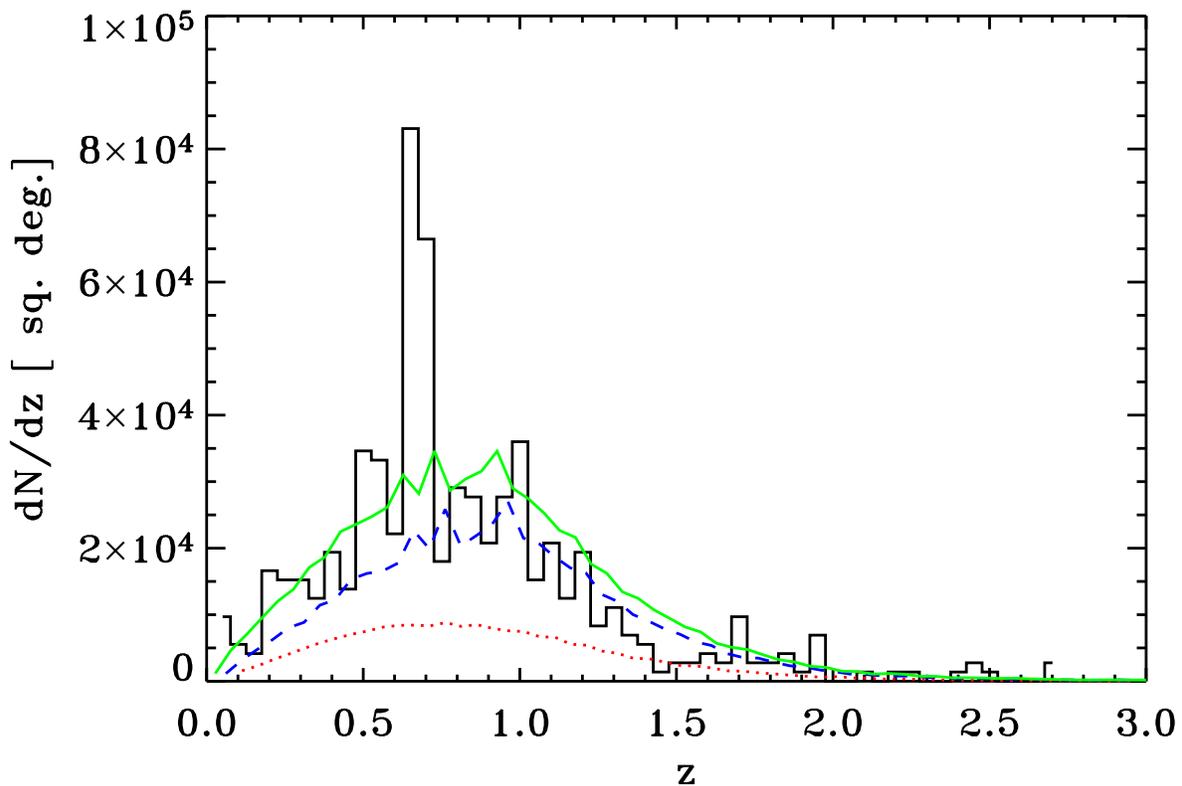}
\caption{The redshift distribution from the K20 sample
  \protect\citep{cimatti2000} compared with the model of
  \protect\citet{xu2003}. The model starbursts are shown in blue
  (middle curve), spheroids in red (lower curve) and total counts in
  green (upper curve).  We have normalized the total counts in the
  model to the total number of sources in the K20 sample (dividing by
  a factor of two).  }\label{fig:nz}
\end{figure}

We know that this model over predicts the 3.6~\micron\ number counts
\citep{lonsdale2004}, so it is important to demonstrate that this model
can nevertheless predict the shape of the redshift distribution.  In
Figure~\ref{fig:nz} we compare the model with the observed redshift
distribution from the K20 sample \citep{cimatti2000}. The shape is a
reasonable fit.  The median redshift of the K20 sample is $z_{\rm
  med}=0.74$; for our model starbursts and spirals we find $z_{\rm
  med}=0.92$, while for the spheroids $z_{\rm med}=0.86$, and the
model as a whole has $z_{\rm med}=0.91$.  The median redshift of the
K20 sample is biased by a sharp peak at $z\simeq0.74$, which is
presumably an artifact of clustering.  So although the K20 sample has
a slightly lower median $z$ than our model, we regard the approximate
agreement between the redshift distributions in Figure~\ref{fig:nz} as
reasonable confirmation of the model.

Using this model we estimate that our $S_{3.6}\ge32$~$\mu$Jy sample
has a median redshift of $z_{\rm med}\approx 0.75$.  If the median
redshift is lower than this, say by a factor of 0.8 as in the K20
comparison above then the median redshift might be $z_{\rm med}\approx
0.6$.

Using the Limber's equation \citep{limber53} inversion technique
adopted by \citet{eduardolss}, which uses a redshift distribution
parameterized by the median redshift, we estimate the real-space
correlation function $\xi(r)=(r/r_0)^{1.8}$ for our combined sample
and find $r_0=4.4\pm0.1h^{-1}$~Mpc. The results for individual fields
are shown in Table~\ref{tab:wtheta}. Adopting the lower redshift,
$z_{\rm med}=0.6$, this drops to $r_0=3.3\pm0.1h^{-1}$~Mpc.  For
comparison the correlation function of quasars in the range
$0.3<z<2.9$ has $r_0=3.99^{+0.34}_{-0.28}h^{-1}$~Mpc, with
$\gamma=1.58^{+0.09}_{-0.10}$ \citep{croom01} and hyper-luminous
infrared galaxies also have a clustering strength similar to AGN at
$z\sim0.7$ \citep{farrah04}.  The uncertainty in our correlation
function is clearly dominated by our uncertainties in the redshift
distribution and emphasizes the need to establish the redshift
distribution of the SWIRE galaxies.


\subsection{Conclusions and future work}

We have performed the first clustering analysis on the SWIRE
survey. We have a very strong detection of clustering with amplitude
similar to $K$-band surveys but with smaller errors. We are thus
consistent with an existing phenomenological model of $K$-band
selected galaxies
$\xi(r,z)=\left(r/r_0\right)^{-\gamma}\left(1+z\right)^{-3}$.
Physically this model could be interpreted as stable clustering,
i.e.\ that galaxies have broken free from the Hubble expansion, though
this seems implausible on large scales. However, since $K$-band
surveys sample different galaxies at different redshifts the
interpretation of the phenomenological model is not this
straightforward.  Now that we have high quality data in the rest-frame
$K$ at high redshift we need phenomenological models specifically
designed to interpret these.

When we have a larger survey area available, and detailed selection
functions, we will be able to extend the analysis to our completeness
limit, which is fainter by a factor of nearly 10 in flux (or 2.5
magnitudes). We will then be able to subdivide our sample by flux and
explore the evolution of clustering in much more detail.  We will
investigate the excess clustering seen on smaller angular scales, and
we will also explore the clustering in the longer wavelength IRAC
bands and the relative clustering of star-forming galaxies and
passively evolving systems, thus gaining insights into the nature of
galaxy bias.

\acknowledgments

We are indebted to Nathan Roche for supplying us with machine readable
versions of his clustering models.  We would like to thank the
referee, Dr.\ Ari Maller, for very useful and constructive comments.
This work was supported by PPARC grant PPA/G/S/2000/00508 (SJO, IW,
EGS).  Support for this work, part of the Spitzer Space Telescope
Legacy Science Program, was provided by NASA through an award issued
by the Jet Propulsion Laboratory, California Institute of Technology
under NASA contract 1407.  This publication makes use of data products
from the Two Micron All Sky Survey
(\url{http://www.ipac.caltech.edu/2mass/}), which is a joint project
of the University of Massachusetts and the Infrared Processing and
Analysis Center/California Institute of Technology, funded by the
National Aeronautics and Space Administration and the National Science
Foundation.

Facilities: \facility{Spitzer Space Telescope}.

\dataset[ads.sa.spitzer#0007770880]{Link to datasets used in this analysis.}
\dataset[ads.sa.spitzer#0007771136]{}
\dataset[ads.sa.spitzer#0005876480]{}
\dataset[ads.sa.spitzer#0005876736]{}
\dataset[ads.sa.spitzer#0005876992]{}
\dataset[ads.sa.spitzer#0005877760]{}
\dataset[ads.sa.spitzer#0005878016]{}
\dataset[ads.sa.spitzer#0005878272]{}
\dataset[ads.sa.spitzer#0005878528]{}
\dataset[ads.sa.spitzer#0005879552]{}
\dataset[ads.sa.spitzer#0005879808]{}
\dataset[ads.sa.spitzer#0005880064]{}
\dataset[ads.sa.spitzer#0008760576]{}
\dataset[ads.sa.spitzer#0008760832]{}
\dataset[ads.sa.spitzer#0008761856]{}
\dataset[ads.sa.spitzer#0008762112]{}
\dataset[ads.sa.spitzer#0008762368]{}
\dataset[ads.sa.spitzer#0008762880]{}
\dataset[ads.sa.spitzer#0008764416]{}
\dataset[ads.sa.spitzer#0008764672]{}
\dataset[ads.sa.spitzer#0008764928]{}
\dataset[ads.sa.spitzer#0008765184]{}

\clearpage

\begin{deluxetable}{lcccccccc}
\tabletypesize{\scriptsize}
\tablecaption{Parameterized $w(\theta)$ 
estimated from each of our sub samples. 
}
\tablewidth{0pt}
\tablehead{
\colhead{Sample} & 
\colhead{Area} &
\colhead{$N_{\rm gals}$} &
\colhead{$N_{\rm stars}$} &
\colhead{$A (\gamma=1.8)$} &
\colhead{$A$} &
\colhead{$\gamma$} &
\colhead{$r_0 (z_{\rm med}=0.75)$} &
\colhead{$r_0 (z_{\rm med}=0.6)$} \\
\colhead{} & 
\colhead{[sq.\ deg.]} & 
\colhead{} & 
\colhead{} & 
\colhead{$[10 ^{-3}]$} & 
\colhead{$[10 ^{-3}]$} &
\colhead{} &
\colhead{$[h^{-1}{\rm Mpc}]$} &
\colhead{$[h^{-1}{\rm Mpc}]$} \\
}
\startdata
Lockman (optical)  &0.3& 2115  & 872 &$2.7\pm0.5$ & $1.2\pm1.3$ & $1.97\pm0.25$ &  $5.5\pm0.5$ &  $ 4.3\pm0.5$  \\
Lockman full tile  &0.4& 3685  & 1551 &$2.0\pm0.3$ & $1.0\pm1.0$ & $1.95\pm0.23$ &  $4.6\pm0.4$ &  $ 3.7\pm0.3$  \\
ELAIS N1 tile\_2\_2&0.8& 8677  & 4163 &$1.7\pm0.2$ & $0.5\pm0.3$ & $2.06\pm0.14$ &  $4.2\pm0.3$ &  $ 3.3\pm0.3$  \\
ELAIS N1 tile\_3\_2&0.8& 8472  & 4272 &$1.7\pm0.2$ & $0.9\pm0.6$ & $1.95\pm0.16$ &  $4.2\pm0.3$ &  $ 3.3\pm0.3$  \\
Combined sample    &2.0& 20834  & 9986 &$1.7\pm0.1$ & $0.6\pm0.3$ & $2.03\pm0.10$ &  $4.4\pm0.1$ &  $3.3\pm0.1$ \\
\enddata
\medskip

Column 3 is the number of galaxies included in each sample, and column
4 is the number of stars that were rejected.  $w(\theta)$ is modeled
as $w(\theta)=A\theta^{1-\gamma}$, with $\theta$ measured in degrees.
Column 5 has $\gamma=1.8$ fixed, while columns 6 and 7 are the results
for a free fit. Columns 8 and 9 are estimates of the strength of the
correlation function $\xi(r)=(r/r_0)^{1.8}$ assuming a median redshift
of $z=0.75$ and $z=0.6$ respectively.  The Lockman (optical) sample is
a subset of the Lockman full tile and is not included in the combined
sample.
\label{tab:wtheta}
\end{deluxetable}

\clearpage

\clearpage

\clearpage

\end{document}